\documentclass[letterpaper, 10 pt, conference]{ieeeconf}  % Comment this line out if you need a4paper

\IEEEoverridecommandlockouts                              % This command is only needed if 
\usepackage{graphics} % for pdf, bitmapped graphics files
\usepackage{epsfig} % for postscript graphics files
\usepackage{times} % assumes new font selection scheme installed
\usepackage{algorithmicx,algpseudocode,algorithm}
\usepackage{amsmath} % assumes amsmath package installed
\usepackage{amssymb}  % For additional symbols
\usepackage{hyperref}
\usepackage{tabularx}
\usepackage{xcolor}
\usepackage{array}
\newtheorem{definition}{Definition}
\newtheorem{example}{Example}

\newtheorem{problem}{Problem}

\usepackage{booktabs} 
\usepackage{subcaption}
\usepackage{multirow}
\usepackage[subtle,tracking=normal]{savetrees}

\usepackage[normalem]{ulem}

\title{\LARGE \bf
NeuroSymbolic Robustness Analysis for \\Discrete Systems with Respect to Transition Deviations}
\author{Shih-Jie Shih$^{1}$ and Jonghan Lim$^{2}$ and Ilya Kovalenko$^{2,3}$, and R\^{o}mulo Meira-G\'{o}es$^{1}$% <-this % stops a space
\thanks{The work of S.J.S. and R.M.G. was supported in part by the U.S. National Science Foundation grant ECCS-2446782.
The work of J.L. and I.K. was supported in part by the U.S. National Science Foundation under CAREER Award No.~2442362.
The authors [1] are with the School of Electrical Engineering and Computer Science, [2] the Department of Industrial and Manufacturing Engineering, and [3] are with the Department of Mechanical Engineering, at The Pennsylvania State University, State College, USA.
{\tt\small \{sfs6681, jxl567, iqk5135, romulo\}@psu.edu}}%
}

\begin{document}

\maketitle
\thispagestyle{empty}
\pagestyle{empty}

%%%%%%%%%%%%%%%%%%%%%%%%%%%%%%%%%%%%%%%%%%%%%%%%%%%%%%%%%%%%%%%%%%%%%%%%%%%%%%%%
\begin{abstract}
Supervisory control of discrete-event systems provides formal guarantees of correctness with respect to a plant model and specification. 
However, these guarantees heavily rely on the plant model, which could deviate from nominal behavior due to modeling errors or faults.
Recent notions of discrete robustness model deviations as a set of additional transitions that are added to the plant.
The discrete robustness is defined as all sets of extra transitions for which the supervised system still guarantees a desired specification.
However, this notion suffers from scalability due to the large solution space and conservatism since most deviations are infeasible in practice.
This paper proposes to address these two issues using a neurosymbolic computing framework for discrete robustness analysis of safety properties. 
First, a neural reasoning layer based on Large Language Models infers a set of feasible deviation transitions from system models, specifications, and domain knowledge. 
Next, a symbolic layer computes the discrete robustness guarantees over the inferred deviation set.
We evaluate our framework on three case studies, demonstrating that our method identifies a smaller set of feasible deviations while preserving robustness guarantees comparable to those of full transition-based analysis.
\end{abstract}
\graphicspath{{Fig/}}

\section{Introduction}
In supervisory control of discrete-event systems, a supervisory controller, or simply supervisor, is synthesized to enforce a desired specification on a plant \cite{Wonham:2018,Lafortune:2021}.
Classical supervisory control and reactive synthesis methods provide a framework for synthesizing supervisory controllers that are correct-by-construction with respect to a given plant model and specification expressed in a formal manner \cite{Ramadge:1987,Pnueli:1989a,Ehlers:2017}.
However, the supervisor's guarantees rely on the plant model, i.e., the specification is only guaranteed for the given plant model. 
In practice, the model of the plant is an approximation of the true system dynamics. 
Mismatches that arise from unmodeled dynamics, sensor noise, unforeseen interactions, or abstraction errors could lead to deviations from the nominal plant model.
These deviations may compromise the supervisor’s ability to guarantee the specification.

In addition to being correct, the ideal supervisor should be robust to \emph{reasonable} plant deviations \cite{meira-goes2023jdeds}.
Several works have investigated robustness notions for discrete systems in supervisory control.
These works investigate the synthesis of supervisors that guarantee the specification even when the plant deviates from its nominal model due to: sensor and actuator faults \cite{Paoli:2011,Rohloff:2012}, model uncertainty \cite{Lin:1993,Young:1995,Cury:1999,Takai:2004}, cyber attacks \cite{Su:2018,meira-goes:2023dealing}, and network communication delays \cite{Alves:2014,Lin:2014}.

Recent work has introduced a new notion of robustness for supervisors and plants modeled as \emph{discrete transition systems} \cite{meira-goes2023jdeds,meira-goes2023cav}.
Deviations are modeled as additional transitions to the nominal plant model. 
In this manner, the notion of discrete robustness of a supervisor is defined as the set of \emph{all} deviations against which the supervisor is robust.

While this new notion of robustness provided new types of system analysis and synthesis tasks, transition-based discrete robustness suffers from two limitations. 
First, due to the combinatorial nature of discrete transition systems, the space of possible transition deviations grows exponentially with the size of the state and action spaces. 
Even for moderately sized systems, enumerating and analyzing all possible deviations becomes computationally intractable as shown in \cite{meira-goes2023jdeds,meira-goes2023cav}. 
Second, the vast majority of deviations considered are \emph{infeasible} in practice: they do not correspond to physically meaningful behavior in the plant.
% As a consequence, the resulting robustness measures are often overly conservative and disconnected from the physical semantics of the system.

In this paper, we address these challenges by introducing a \emph{neurosymbolic} computing framework for discrete robustness analysis of safety properties. 
Neurosymbolic computing combines the advances of learning in neural networks with the reasoning abilities of symbolic methods \cite{garcez2019neural}.
In our context, we decouple the deviation feasibility reasoning from formal robustness computation. 
Using Large Language Models (LLMs) as a neural reasoning layer, we automatically infer a subset of physically feasible deviation transitions from the plant and supervisor models, natural-language system descriptions, a formal specification model, and domain knowledge. 
This neural component identifies a set of \emph{feasible transition deviations} that are consistent with the semantics of the system.

On top of this feasible transition deviation set, we apply a symbolic robustness analysis that computes discrete robustness guarantees with respect to safety properties. 
Since the symbolic layer operates on a formally defined transition system and a well-specified deviation model, the resulting robustness guarantees remain correct. 
In this manner, our approach combines the scalability and semantic reasoning capabilities of neural models with the correctness guarantees of symbolic methods.
% To the best of our knowledge, this paper is the first one to study a neurosymbolic approach to bridge the gap between scalability and correctness guarantees in discrete event systems. 
% To the best of our knowledge, this paper is the first one to study a neurosymbolic approach in discrete event systems. 

We have built a prototype of the proposed neurosymbolic framework for computing robustness and applied it to three case studies: (1) a manufacturing system, (2) a communication protocol, and (3) a radiation therapy interface. 
Our results demonstrate that the proposed approach is able to identify a smaller set of physically feasible deviations, while providing robustness guarantees that are comparable to those obtained through the full transition-based analysis.

The contributions of this paper are as follows: (i) A neurosymbolic framework that leverages LLMs to infer physically feasible deviation models and a symbolic method to compute robustness guarantees with respect to the inferred deviation set,
(ii) A prototype tool for computing robustness using our neurosymbolic framework and an experimental evaluation on three case studies.

The rest of this paper is organized as follows. 
We provide a review of labeled transition systems (LTS), supervisory control, and discrete robustness in Sect.~\ref{sect:prelim}.
In Sect.~\ref{sect:problem}, we described the problem of computing neurosymbolic discrete robustness of supervisors.
Our neurosymbolic framework to compute robustness is presented in Sect.~\ref{sect:solution}.
Section~\ref{sect:case-studies} evaluates our framework in three case studies.
We conclude this paper in Sect.~\ref{sect:conclusion}.

\section{Preliminaries} \label{sect:prelim}
\subsection{Labeled Transition Systems}
In this work, we model the behavior of dynamical systems using finite labeled transition systems (LTS) \cite{Baier:2008}.
Formally, an LTS is defined as follows:
\begin{definition} 
A \emph{labeled transition system} (LTS) $E$ is a tuple $\langle Q_E, \text{Act}_E, R_E, q_{0,E}\rangle$, where $Q_E$ is a finite set of states, $\text{Act}_E$ is a finite set of actions, $R_E\subseteq Q_E\times \text{Act}_E \times Q_E$ is the transition relation of $E$, and $q_{0,E} \in Q_E$ is the initial state. 
\end{definition}
We extend the transition relation $R_E$ to finite sequences of actions as $R^*_E \subseteq Q_E \times \text{Act}^*_E \times Q_E$. 
A \emph{trace} of LTS $E$ is a finite sequence of actions $a_0 a_1 \cdots a_n \in \text{Act}_E^*$ such that $(q_{0,E}, a_0 a_1 \cdots a_n, q) \in R_E^*$ for some $q \in Q_E$. 
The set of all traces of $E$ is denoted $\mathrm{beh}(E)$. 
% Actions are either controllable or uncontrollable. 
An LTS is \emph{deterministic} if for any $(q, a, q'), (q, a, q'') \in R$, then $q' = q''$; otherwise it is \emph{nondeterministic}. 
% In $(q, a, q')$, $q$ is the predecessor state, $q'$ is the successor state. 
Given two LTSs $E_1$ and $E_2$, their parallel composition $E_1 \| E_2 = \langle Q_{E_1} \times Q_{E_2}, \text{Act}_{E_1} \cup \text{Act}_{E_2}, R_{E_1 \| E_2}, (q_{{0}_{E_1}}, q_{{0}_{E_2}}) \rangle$ synchronizes over common actions and interleaves the remaining actions~\cite{Lafortune:2021,Baier:2008}.

\subsection{Supervisor}

Given an LTS modeling the uncontrolled plant, a supervisor is a control mechanism that enforces properties by disabling actions~\cite{Ramadge:1987}.
Due to limited actuation capabilities, we partition the set of actions into controllable $Act_{c}$ and uncontrollable $Act_{uc}$ actions. 
A supervisor only disables controllable actions.
Formally, we define a supervisor as:
\begin{definition}
% Let an LTS $E$ represent the system model to be controlled. 
A supervisor for $E$ is a function $S: Act^* \to \Gamma$ that maps a finite sequence of actions to a set of admissible actions $\Gamma = \{\gamma \subseteq Act \mid Act_{uc}\subseteq \gamma\}$.
\end{definition}

A \emph{controlled trace} of $E$ is a trace $a_0 \ldots a_n \in \text{beh}(E)$ such that $a_i \in S(a_0 \ldots a_{i-1})$ for $i \leq n$. 
The set of all controlled traces, denoted by $\text{beh}(E/S)$, defines the \emph{closed-loop system} of $S$ controlling $E$, denoted $E/S$.
% For $E/S$, the supervisor $S$ can only disable controllable actions, while uncontrollable actions cannot be disabled by the supervisor.
We assume that supervisor $S$ has finite memory and is represented by a deterministic LTS. 
With an abuse of notation, the supervisor $S$ is also denoted by $S = \langle Q_S, \text{Act}_S, R_S, q_{0,{S}} \rangle$ where $\text{Act}_S = \text{Act}_E$. 
In this manner, the closed-loop system is represented by the composition of $E$ and $S$: $E/S = E \| S$ 

\subsection{Safety Properties}
Safety properties are regular linear-time properties of system behaviors that specify that ``something bad never happens''~\cite{Baier:2008}.
A \emph{safety property} $P$ over a system $E$ is represented by a deterministic LTS $P$ that defines the set of accepted behaviors. 
The LTS $P$ encodes both traces that satisfy $P$ and those that violate it by including a sink error state $\text{err} \in Q_P$. An LTS $E$ satisfies property $P$, denoted $E \models P$, if and only if the traces in $\text{beh}(E)$ do not reach the error state $\text{err} \in Q_P$.
We test if $E \models P$ by composing $E \| P$ and checking whether $\text{err}$ is reachable.

\begin{example}
\label{eg: manufacturing}
Consider a small manufacturing cell $G$ consisting of two components: (1) a machine $M_1$ that processes a part at a time (Fig.~\ref{fig:machine-buffer:M1}), and (2) a buffer $B$ with unit capacity that temporarily stores parts after processing (Fig.~\ref{fig:machine-buffer:B}). 
The model for the manufacturing cell is the composition of them: $G=M_1\|B$.
Considering that $Act_c = \{in_1,out\}$, supervisor $S$ in Fig.~\ref{fig:machine-buffer:S} guarantees that the buffer does not overflow, i.e., state $OF$ is not reached.

\begin{figure}[thpb]
\begin{subfigure}[t]{0.35\columnwidth}
\centering
\includegraphics[width=1\columnwidth]{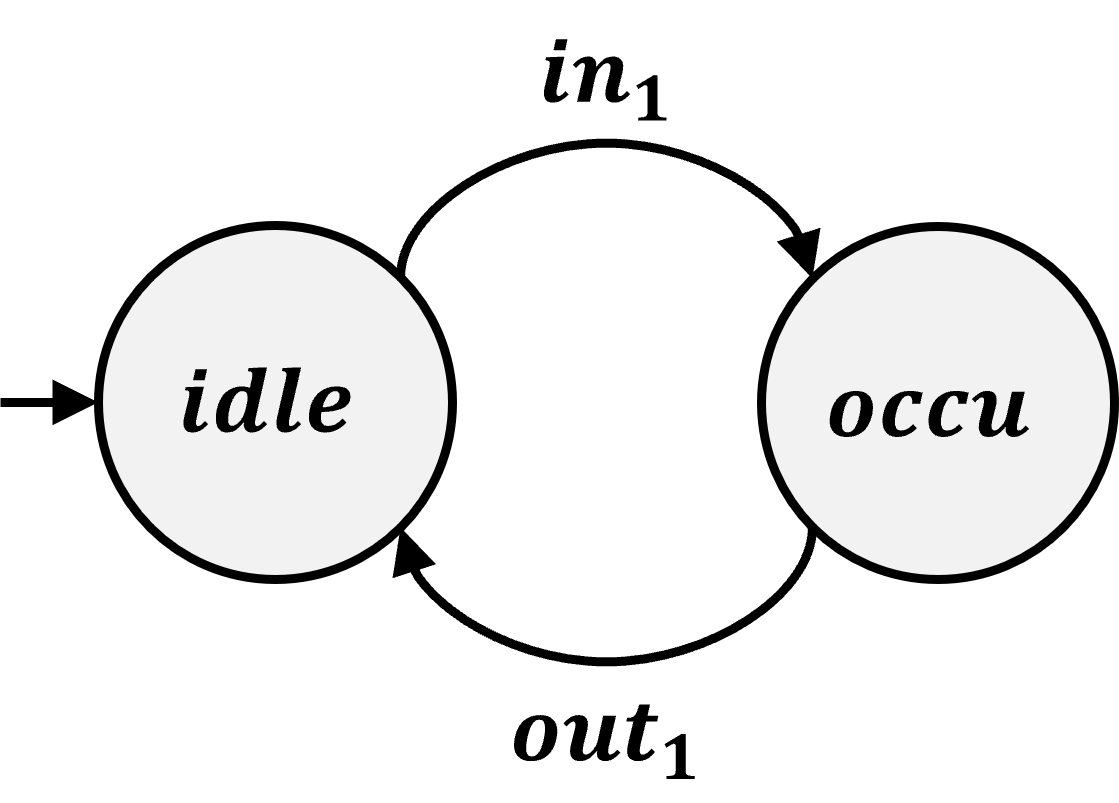}
\caption{Machine $M_1$}
\label{fig:machine-buffer:M1}
\end{subfigure}
\ 
\begin{subfigure}[t]{0.6\columnwidth}
\centering
\includegraphics[width=1\columnwidth]{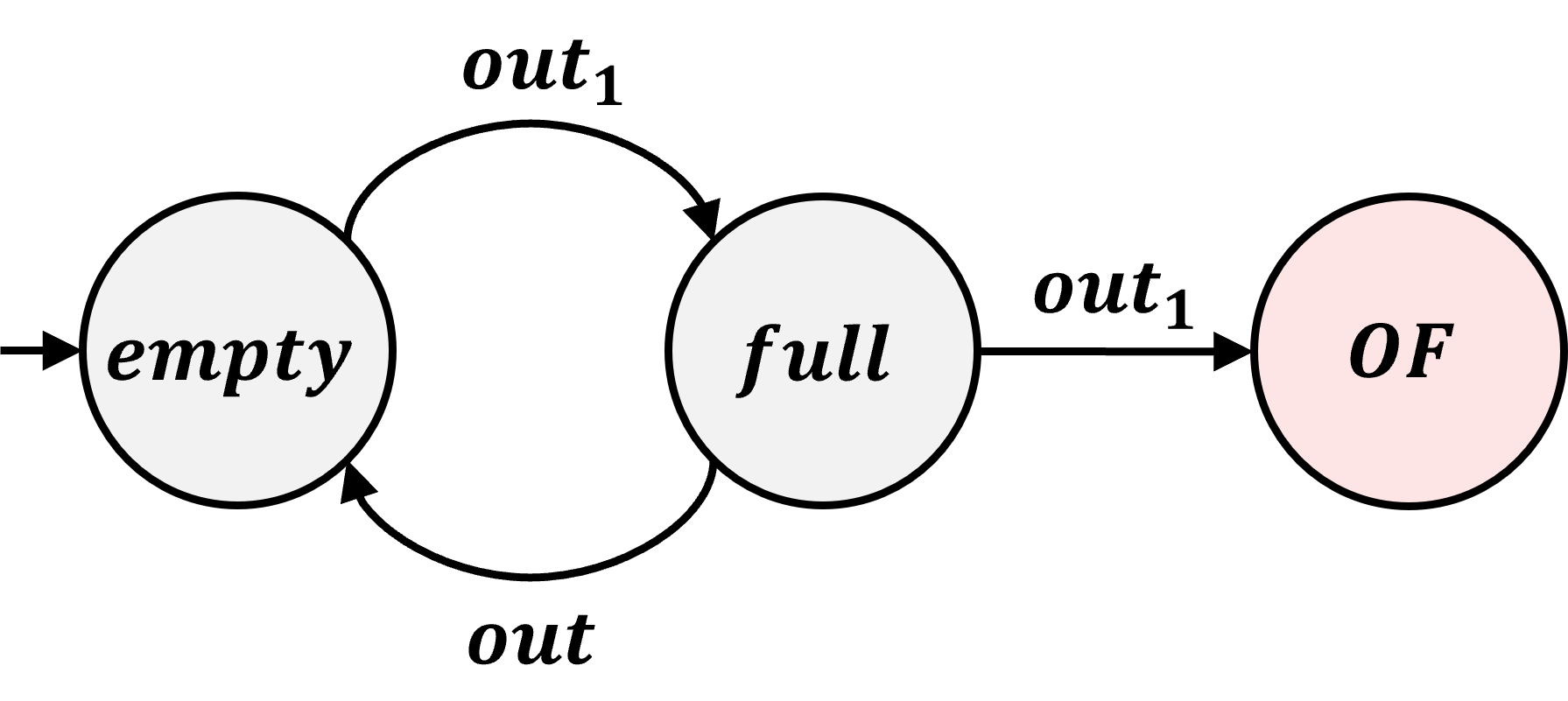}
\caption{Buffer $B$}
\label{fig:machine-buffer:B}
\end{subfigure}
\ 
\begin{subfigure}[t]{1\columnwidth}
\centering
\includegraphics[width=0.5\columnwidth]{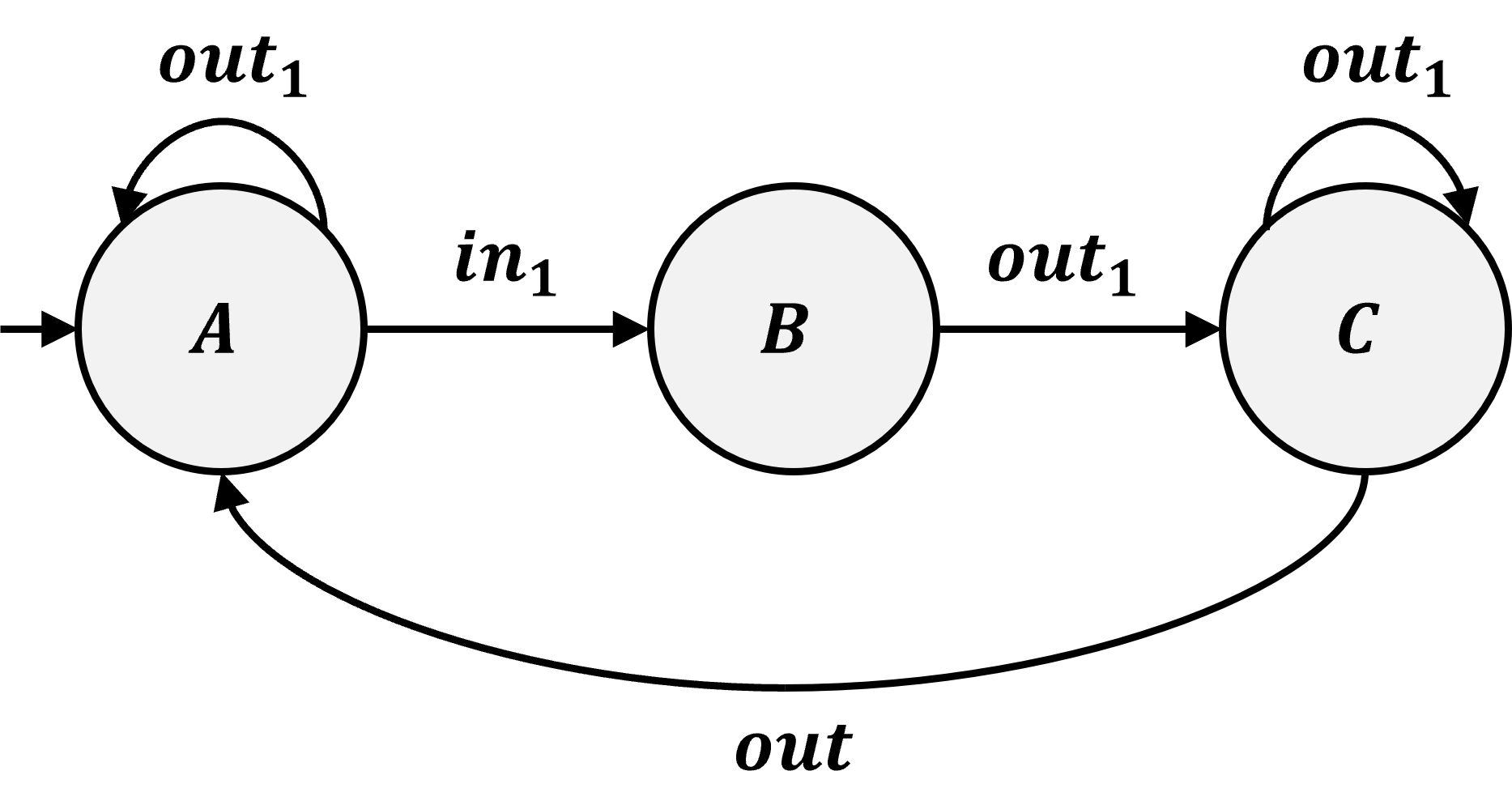}
\caption{Supervisor $S$}
\label{fig:machine-buffer:S}
\end{subfigure}
\caption{Models of plant and supervisor for manufacturing example}
\label{fig:automation_system}
\vspace{-1em}
\end{figure} 
\end{example}

\subsection{Robustness}
% \subsection{Deviations}
To account for mismatches between the model of the system and the true system, deviations are modeled as additional transitions to LTS $E$.
A \emph{deviation} is a set of transitions $d \subseteq  (Q_E \times \text{Act}_E \times Q_E)$.
\begin{definition}
Given an LTS $E$ and a deviation $d$, we define the \emph{deviated system} $E_d$ as:
\begin{equation}
E_d := \langle Q_E, \text{Act}_E, R_E \cup d, q_{0_{E}} \rangle
\end{equation}
\end{definition}

For LTS $E$, a supervisor $S$ guarantees property $P$, i.e., $E/S \models P$. For the deviated system $E_d$, with the same supervisor, it might violate this property, i.e., $E_d/S \not\models P$.
Robustness captures the ability of a controlled system to maintain its safety properties under system deviations~\cite{meira-goes2023cav}. 
\begin{definition}[Robust Supervisor]
Supervisor $S$ is a \emph{robust supervisor} with respect to system $E$, deviation $d$, and property $P$ if $E_d/S \models P$.
\end{definition}
For convenience, we say that a deviation is \emph{robust} with respect to $E$, $S$, and $P$ if $S$ is robust to $d$.
% \begin{definition}[Robust Deviation]
% Deviation $d$ is a \emph{robust deviation} with respect to $E$, $S$, and $P$ if $S$ is a robust supervisor with respect to $E$, $d$, and $P$.
% \end{definition}
Based on this deviation model, the discrete robustness $\Delta$ is defined as in \cite{meira-goes2023cav}.
\begin{definition}[Robustness]
\label{def:robustness}
Let system $E$, supervisor $S$, property $P$ such that $E/S \models P$, and system constraint $P_{\text{sys}}$ such that $E \models P_{\text{sys}}$ be given. 
The \emph{robustness} of supervisor $S$ with respect to $E$, $P$, and $P_{\text{sys}}$, denoted by $\Delta(E,S,P,P_{\text{sys}})$, is a set of robust deviations $\Delta \subseteq 2^{Q_E \times \text{Act}_E \times Q_E}$ satisfying:
\begin{enumerate}
    \item $\forall d \in \Delta$: $E_d/S \models P$ \hfill [d is robust]
    \item $\forall d \subseteq Q_E \times \text{Act}_E \times Q_E$: $(E_d/S \models P \wedge E_d \models P_{\text{env}}) \Rightarrow \exists d' \in \Delta: d \subseteq d'$ \hfill [d is represented]
    \item $\forall d, d' \in \Delta$: $d \neq d' \Rightarrow d \not\subseteq d'$ \hfill [unique representation]
    \item $\forall d \in \Delta$: $E_d \models P_{\text{sys}}$ \hfill [d is feasible]
\end{enumerate}
\end{definition}
When $E$, $S$, $P$, and $P_{\text{sys}}$ are clear from context, we simply write $\Delta$. 
The set $\Delta$ defines an upper bound on the possible deviations from $E$ that supervisor $S$ is robust against, i.e., the closed-loop system remains safe. 
The system constraint $P_{\text{sys}}$ captures domain knowledge about the system under analysis, filtering environment deviations that might not be physically \emph{feasible} to analyze.

% By definition, $\Delta$ is always non-empty since $d = \emptyset$ is always robust. 
% Moreover, due to conditions (2) and (3), only maximal robust deviations are included in $\Delta$.

% \rmg{If we have time - add example talking about robustness of our running example.}

\section{Problem Statement} \label{sect:problem}
% For feedback - you are repeating the same information above.
% This is the Problem Statement section, but you have not defined a problem. 

% In \cite{meira-goes2023cav}, deviations are modeled as additional transitions
% $d \subseteq (Q_E \times \text{Act}_E \times Q_E)$ that augment the original system E, and robustness is defined as the maximal set of
% deviation sets for which the safety property P is preserved. However, a few issues arise with this approach.
Computing $\Delta$ over all possible transition deviations, over $2^{|(Q_E \times \text{Act}_E \times Q_E) \setminus R_E|}$, has the benefit of covering every possible deviation.
However, there are two major challenges with its computation and analysis: (i) the scalability of computing $\Delta$, and (ii) the connection between model deviations and feasible deviations in the real system.

% For a system $E = \langle Q_E, \text{Act}_E, R_E, q_{0_{E}} \rangle$, 
% the set of all possible deviations is $ (Q_E \times \text{Act}_E \times Q_E) \setminus R_E$, representing all transitions that do not currently exist in the system. 
Since a deviation $d$ is any subset of 
$ (Q_E \times \text{Act}_E 
\times Q_E) \setminus R_E$, the total number of possible deviation sets grows exponentially with the system size. 
Our small manufacturing running example has $2^{6\times 3\times 6 - 7} = 2^{101}$ possible deviations.
Although limiting the property to invariance in \cite{meira-goes2023jdeds} and using heuristics in \cite{meira-goes2023cav} improved the scalability in comparison to exhaustive methods, results could be extended to only moderately-sized systems.

Moreover, treating every possible deviation as feasible introduces a fundamental mismatch between the formal model and physical system. 
% The set $\mathcal{D}$ includes transitions that are purely mathematical constructs without physical meaning. 
For instance, in Example~\ref{eg: manufacturing}, adding a transition $(\text{Idle},\ out_1,\ \text{Idle})$ to Machine~1's model in Fig.~\ref{fig:machine-buffer:M1} implies that Machine~1 outputs parts spontaneously, regardless of whether a part has arrived in the system.
Although this deviation is mathematically possible, it is infeasible in the real system.
% would force the buffer a transition representing a part spontaneously appearing in an idle machine has no physical basis despite being mathematically valid. 
% Similarly, mechanical constraints may prohibit certain state transitions, and system design may explicitly rule out certain sequences of events. 
Including these physically infeasible deviations not only inflates the search space but also produces robustness results that do not accurately reflect real-world system behavior.
Therefore, after computing $\Delta$, the engineer would need to evaluate if certain robust deviations are feasible or not.

In \cite{meira-goes2023cav}, the property $P_{sys}$ reduces the number of infeasible transitions, i.e., only transitions that satisfy $P_{sys}$ are considered as in Def.~\ref{def:robustness}.
% However, the addition of $P_{sys}$ increases the complexity of the problem since it introduces a new model checking task. 
However, adding $P_{sys}$ introduces an additional model-checking step to the algorithmic pipeline, increasing the overhead on computing $\Delta$.
Moreover, defining $P_{sys}$ requires detailed specifications and domain knowledge from domain experts.
Extracting and formalizing this knowledge places a burden on these stakeholders, requiring them to explicitly encode physical laws, mechanical limitations, and design intent into formal specifications.

% Based on the above mentioned issues, we need a method to compute $\Delta$ that (i) considers only feasible deviations $D\subseteq Q_E\times \text{Act}_E\times Q_E$; (ii) is scalable.
To address these issues, we consider limiting the computation of $\Delta$ to only consider a subset of the possible deviations $D\subseteq Q_E\times Act_E \times Q_E$.
\begin{definition}
\label{def:feas_rob}
Let system $E$, supervisor $S$, property $P$ such that $E/S \models P$, and a set of feasible deviations $D\subseteq Q_E\times Act_E \times Q_E$ such that $E \models P_{\text{sys}}$ be given. 
The \emph{robustness} of supervisor $S$ with respect to $E$, $P$, and $D$, denoted by $\Delta_f(E,S,P,D)$, is a set of robust deviations $\Delta_f \subseteq 2^{D}$ satisfying:
\begin{enumerate}
    \item $\forall d \in \Delta_f$: $E_d/S \models P$
    \item $\forall d \subseteq D$: $(E_d/S \models P) \Rightarrow \exists d' \in \Delta_f: d \subseteq d'$
    \item $\forall d, d' \in \Delta_f$: $d \neq d' \Rightarrow d \not\subseteq d'$
\end{enumerate}
\end{definition}

Definition~\ref{def:feas_rob} is defined similarly as Def.~\ref{def:robustness}, the main difference is using the set of feasible deviation transitions $D$ instead of a property $P_{sys}$.
Based on Def.~\ref{def:feas_rob}, we define the problem of computing robustness. 
\begin{problem}\label{prob:feasible}
Given system $E$, supervisor $S$, property $P$ such that $E/S \models P$ and a set of feasible deviations $D\subseteq Q_E\times Act_E \times Q_E$.
Compute robustness $\Delta_{f} \subseteq 2^D$ with respect to $E$, $S$, $P$, and $D$.
\end{problem}

The problem of finding $\Delta_f$ remains similar to computing $\Delta$.
When $D$ is large, this problem also suffers from a large solution state space.
Therefore, one of the main challenges is finding a small set of deviations $D$.
We assume that the set of \emph{feasible deviations}, those that are physically possible in the plant, is relatively small compared to the set of all possible deviations. 

% \textbf{Issue 1 (State-Space Explosion):} The number of possible deviation sets grows as $2^{|(Q_E \times \text{Act}_E \times Q_E) \setminus R_E|}$, making exhaustive enumeration computationally difficult for practical systems.

% \textbf{Issue 2 (Infeasible Deviations):} The mathematical set $Q_E \times \text{Act}_E \times Q_E$ includes transitions that are not physically realizable, leading to an artificially inflated search space and potentially misleading robustness analysis.

% \textbf{Issue 3 (Knowledge Elicitation):} Constructing a subset of physically feasible deviations requires detailed system specifications and domain knowledge from designers and engineers, creating a significant manual effort barrier.
\begin{figure*}[t]
    \centering
    \includegraphics[width=0.95\textwidth]{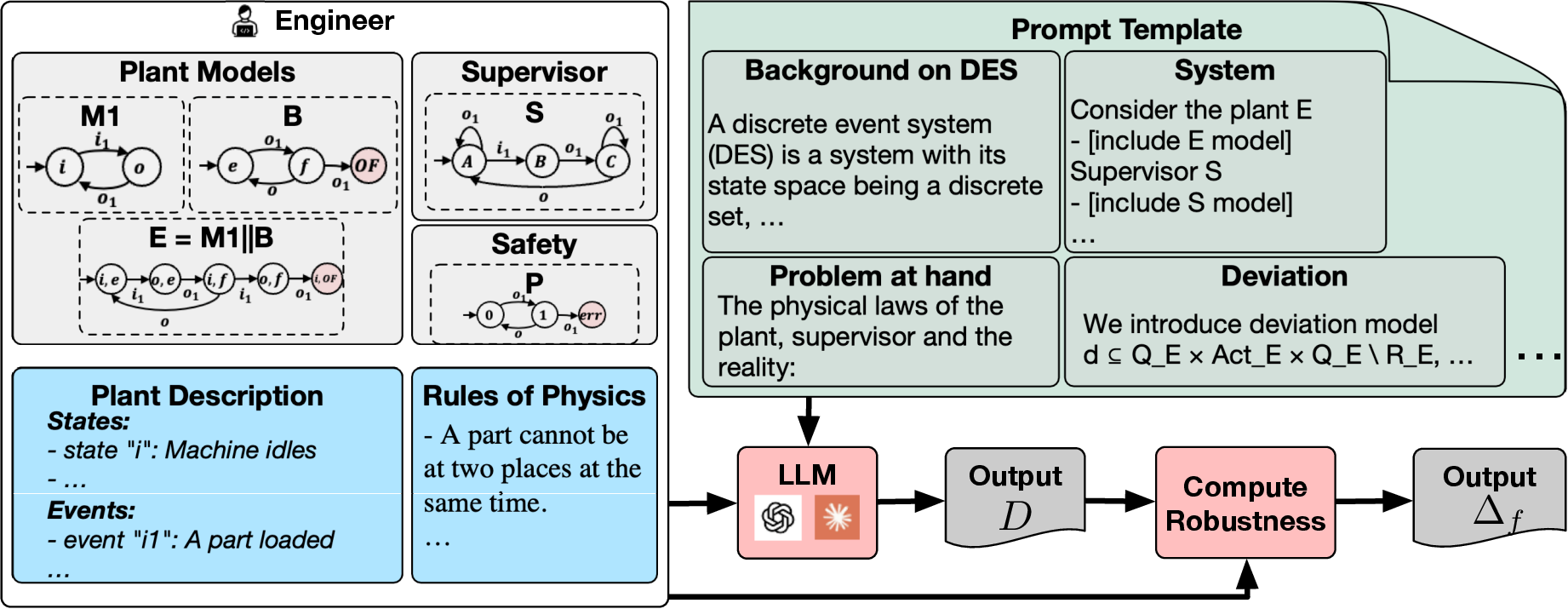}
    \caption{Overview of our neurosymbolic approach to compute robustness.}
    \label{fig:diagram}
    \vspace{-1em}
\end{figure*}

\section{Solution Methodology}\label{sect:solution} 
% \rmg{We got to work on this section to clarify our solution. 
% I will work on this tomorrow morning.}

As mentioned in Section~\ref{sect:problem}, one of the main challenges with Problem~\ref{prob:feasible} is finding the set of deviations $D$.
In this section, we introduce our neurosymbolic framework as a solution to Problem~\ref{prob:feasible}.
Figure~\ref{fig:diagram} provides an overview of our solution framework.

First, we leverage Large Language Models (LLMs) to address the issue of knowledge elicitation by extracting domain knowledge through natural language and formal models.
LLMs provide capabilities for natural language understanding and generation \cite{bubeck2023sparks}.
For instance, LLM models such as OpenAI’s GPT-4 \cite{achiam2023gpt} and Google's BERT \cite{devlin2019bert} introduce new opportunities to extract formal models from natural language descriptions.
We use LLMs to construct a set of feasible deviations $D\subseteq Q_E\times Act_E\times Q_E$.
% More specifically, we provide a prompt to an LLM based on the system models and natural language specification.

After a set of feasible deviations is obtained, we focus on computing $\Delta_f$.
% One approach to compute $\Delta_f$ is to also use LLMs.
% However, LLMs cannot guarantee that all conditions in Def.~\ref{def:feas_rob} are met when computing $\Delta_f$.
Given that the conditions in Def.~\ref{def:feas_rob} are symbolic conditions (model checking and set conditions), we then use symbolic reasoning to compute them.
In other words, we provide an algorithm that computes $\Delta_f$ ensuring that all conditions in Def.~\ref{def:feas_rob} are satisfied.
% This neural component then helps mitigate by filtering out physically unrealizable transitions based on real-world constraints. 
% Next, once the LLM provides the set of feasible deviations, we can search for robustness
% 
% Finally, the reduced search space partially mitigated the issue of state-space explosion, though symbolic verification is still required for the remaining candidates.

\subsection{Overview Neurosymbolic Method}
Our neurosymbolic approach combines the reasoning capabilities of LLMs with the formal guarantees of symbolic model checking. 
The methodology consists of two stages as shown in Fig.~\ref{fig:diagram}:

\noindent
\textbf{Stage 0: LTS Models, Language Description, and Prompt Template.}
First, we collect the system's information from the engineer.
We assume that the models of the environment $E = E_1||E_2||\dots||E_n$ are given in the format of $n$ LTS components.
Descriptions of states and events are provided for each component.
In addition, a set of natural language physical descriptions of the system is also provided.
The engineer also provides the LTS models of the supervisor and safety specification.
To prompt the LLM, we use a prompt template that integrates the system models with the natural-language description defined above.

\noindent
\textbf{Stage 1: Neural Feasibility Filtering.}  
We use an LLM to reason about physical constraints expressed in natural language and domain knowledge from the LTS model of the plant. 
Given system descriptions and constraint specifications, the LLM identifies which potential deviations are physically realizable, producing a list of feasible deviations ${D} \subseteq 2^{(Q_E \times \text{Act}_E \times Q_E) \setminus R_E}$.

\noindent
\textbf{Stage 2: Symbolic Robustness Verification.} 
We provide a brute-force algorithm for our robustness computation.
For each candidate deviation $d \subseteq D$, we verify whether $E_d/S \models P$. 
% This stage provides formal guarantees that the identified robust deviations truly satisfy the safety property.order 
% To construct $\Delta_f$, we extract
% This stage provides formal guarantees that the identified robust deviations truly satisfy the safety property.order 
To construct $\Delta_f$, we extract only the maximal robust deviations according to the subset ordering.
This step ensures that $\Delta_f$ contains no redundant elements and satisfies the conditions in Def.~\ref{def:robustness}.

% The key insight of this approach is that the LLM reduces the search space from $2^{|(Q_E \times \text{Act}_E \times Q_E) \setminus R_E|}$ to $|{D}|$, which is typically several orders of magnitude smaller. 
% While the symbolic verification stage still has exponential worst-case complexity, it operates on a reduced input size.
%Next, we provide details on the prompt template and the algorithm for our robustness computation.

\subsection{Prompt Template Construction}
% The LLM-based feasibility selection addresses the knowledge elicitation challenge, albeit with some constraints, in both natural language and the formal plant description model. 
% This stage takes as input: (1) the description of the system in the LTS model, (2) the physical constraints and domain rules, and (3) the set of all possible deviations.

% \textbf{Prompt Construction.} 
We construct a structured prompt template that provides the LLM with (1) background knowledge on LTS and supervisory control, (2) LTS models of plant, supervisor, components of them (3) meaning of the states, actions, (4) safety property model, (5) natural language of physical laws, (6) definition of deviation, and (7) optional examples to demonstrate a physically infeasible deviations. 
Part of this template information is shown in Fig.~\ref{fig:diagram}.
In the prompt template, we also provide the following natural language objective:
\begin{quote}
Generate ALL feasible deviations representing component faults. Deviations must:
(1) Be minimal (no redundant transitions); (2) Have physical meaning (represent realistic faults);(3) Respect physical laws listed above.
\end{quote}
Lastly, the template asks for the output in the following format:
\begin{quote}
Organize deviations by state. For each deviation, provide - Deviation: $s_0 --- e ---> s_1$
\end{quote}
% \textbf{LLM Reasoning.} 
% The LLM analyzes whether the deviation violates any physical constraint. 
% For example, given the deviation that a part leaves an empty machine $M_1$, the LLM identifies that a part cannot leave an idle machine (violates conservation of mass), classifying it as infeasible. 
% % In this work, we use OpenAI ChatGPT and Anthropic Claude.

% \textbf{Logic of Reasoning.} 
% To improve accuracy, we provided the mathematical definition of deviation, prompting LLM to look for all possible added transitions, including self-loops or nondeterministic transitions.

\subsection{Robustness Computation}

As discussed in \cite{meira-goes2023cav}, one approach to solving Problem.~\ref{prob:feasible} is a brute-force algorithm as shown in Alg.~\ref{alg:rob}. 
Algorithm~\ref{alg:rob} proceeds in two stages: (i) computing the set of robust deviations from $D$, and (ii) identifying the maximal elements of this set. 
In the first stage, we verify $E_d /S \models P$ for all possible feasible deviations $d\subseteq D$ using standard model-checking techniques \cite{Baier:2008}.
Next, we compute $\Delta_f$ by removing non-maximal elements.
Since the algorithm exhaustively checks every candidate deviation set for robustness, it follows that it computes $\Delta_f$.

\begin{algorithm}
\caption{Compute-Robustness}
\label{alg:rob}
\begin{center}
\renewcommand{\algorithmicrequire}{\textbf{Input:}}
\renewcommand{\algorithmicensure}{\textbf{Output:}}
\begin{algorithmic}[1]
\Require{$E$, $S$, $P$ and $D$}
\Ensure{$\Delta_f$}
\State $\Delta_f := \emptyset$
\ForAll{$d\subseteq D$}
\If{$E_d/S\models P$}
\State $\Delta_f := \Delta_f\cup \{d\}$
\EndIf
\EndFor
\While{$\exists d_1,d_2\in\Delta_f$ s.t. $d_1\subset d_2$} %Algorithmic (double loop) forall 
\State $\Delta_f := \Delta_f\setminus \{d_1\}$
\EndWhile
\State \Return $\Delta_f$ 
% \vspace{-0.5em}
\end{algorithmic}
\end{center}
% \vspace{-1.5em}
\end{algorithm}

\section{Case Studies}\label{sect:case-studies}

\subsection{Case Study Setup}
We implemented our \textit{neurosymbolic} framework using Anthropic's Sonnet 4.5 (Claude) and OpenAI's GPT-5.2 (GPT) for feasibility filtering. 
% The LLMs are prompted to enumerate all possible transitions, including self-loops and nondeterministic transitions, excluding transitions in the nominal plant.
% The symbolic verification of Alg.~\ref{alg:rob} was implemented in Python using the MDESops tool \cite{meira2023mdesops}.
We implemented the symbolic verification of Alg.~\ref{alg:rob} in Python using MDESops \cite{meira2023mdesops}.
For convenience, we discuss each case study below using the set of feasible deviations from GPT. 
In Section~\ref{sect:disc}, we summarize the results from both Claude and GPT as well as discuss some differences between them. 
We ran our algorithm on a Dell XPS 17 with an i7-13700H CPU and 32 GB RAM running Ubuntu 22.04 LTS.
% Before the symbolic verification, feasible and infeasible deviations, along with the reasons behind, are checked.

% \rmg{Add a subsection here before everything talking about the setup. For example, which LLMs we used and which models.}

% \rmg{Overall comment regarding the case study. To me, it is more interesting to tell what we uncover is feasible under our perspective (engineer/designer).
% Right now, you do not cover this. You just mention numbers, which does not showcase why people should get excited about this methodology.
% What are some of the deviations you found in each of the case studies that you can reason they are feasible? 
% I am talking about one to three(max) deviations. 
% We had many meetings where we discussed these deviations and their "feasibility". 
% To be fair, you can also say that we found some that we think are infeasible. \\
% Next, show that because these feasible deviations (the ones you talked based on my comment above) are there we have interesting robustness results. Some of them are robust and some are not. 
% Ideally, we can say that we find the same robustness as in my CAV paper. 
% See my CAV paper for how in each case study we discuss the robustness we found. We talk about specific deviations that the system is robust and some that are not.
% }

\subsection{Manufacturing}
\noindent
\textbf{Overview} In Sect.~\ref{sect:prelim}, we introduced the small manufacturing example with models depicted in Fig.~\ref{fig:automation_system}.
Herein, we analyze the robustness of this system using our framework.

\noindent
\textbf{Calculating Robustness} 
% The manufacturing system has $|Q|=\text{6}$, $|Act|=\text{3}$, $|R|=\text{7}$, resulting in 101 possible additional transitions. The LLM analyzed the system using physical constraints, including that parts cannot spontaneously appear or disappear, the machine processes one part at a time, and the buffer has unit capacity. 
% \textit{Claude} identified 62 possible deviations representing sensor failures, actuator failures, and component losses. 
% Symbolic verification found that 52 deviations (83.9\%) maintain safety while 10 deviations (16.1\%) cause violations of the safety property.
We used our tool to calculate the robustness of this manufacturing example. 
In this case, GPT identified $|D|=18$ feasible deviations.
Based on $D$, we computed the robustness $\Delta_f$, finding a sole maximal robust deviation.
This result agrees with Theorem~1 in \cite{meira-goes2023jdeds}, which states that for invariance properties, a unique robust deviation exists.

\begin{figure}[t]
    \centering
    \includegraphics[width=0.7\columnwidth]{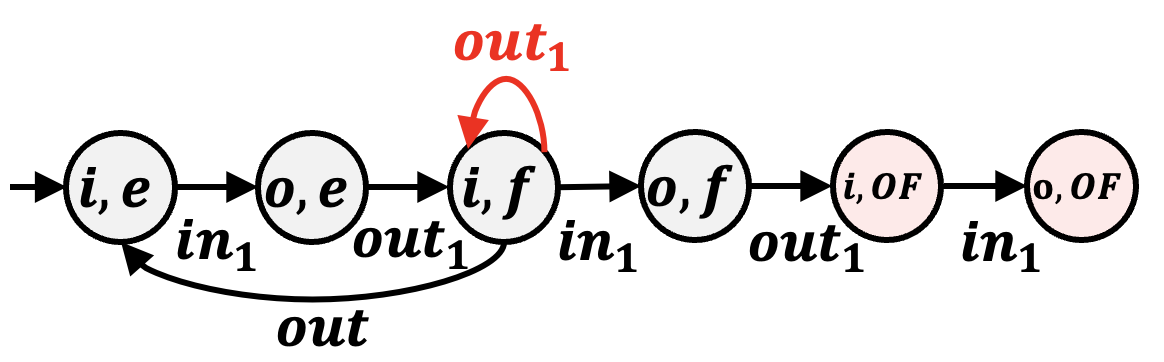}
    \caption{A not robust deviated manufacturing plant. Transition in red is added to the nominal plant.}
    \label{fig:mfg-not-robust}
    \vspace{-1em}
\end{figure}

This maximal deviation reports that the manufacturing system is robust against $15$ transitions.
For instance, it is robust to a faulty loading of a part in Machine 1, transition $('idle', 'empty'),\allowdisplaybreaks in1, \allowdisplaybreaks ('idle', 'empty')$.
On the other hand, the supervisor is not robust to a faulty unloading of the part from the buffer as depicted in Fig.~\ref{fig:mfg-not-robust}.
In this case, the buffer attempts to unload a part via event $out$, but the part remains in the buffer. 
Because the supervisor assumes that the unload operation is successful, it subsequently allows a new part to enter the system, which leads to a buffer overflow.

\subsection{Therac-25}
\begin{figure}[t]
    \centering
    \includegraphics[width=0.5\columnwidth]{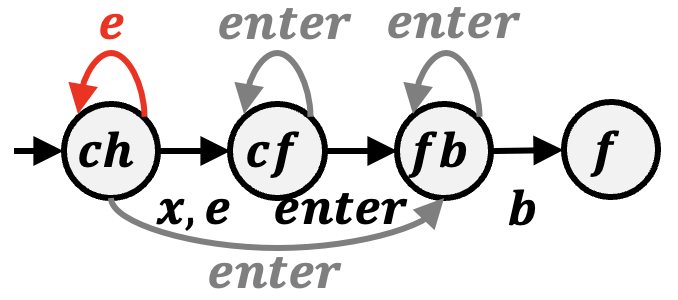}
    \caption{Plant model for Therac-25. Transitions in black are part of the nominal model, transitions in gray are robust, while the red transition is not a robust deviation.}
    \label{fig:therac}
    % \vspace{-2em}
\end{figure}

\noindent
\textbf{Overview} Therac-25 is a radiation therapy machine with a record of causing radiation overdose to patients, some leading to fatality due to a design flaw \cite{leveson1993investigation}. 
We followed the LTS models of the Therac-25 from \cite{meira-goes2023cav}.
The environment $E$ models the operator workflow of a nurse during radiation treatment administration as shown in Fig. ~\ref{fig:therac}.
The model consists of four sequential stages: $ch$ for choose mode, $cf$ for confirm mode, $fb$ for fire beam, and $f$ for finished. 
Operators select beam mode by selecting X-ray ($x$) or electron therapy ($e$), confirm the selection ($enter$), and then initiate radiation delivery ($b$).
We modeled the supervisor as the composition of three components: (1) a computer terminal, (2) a turntable, and (3) a beam emitter. 
The safety property is defined by ``X-ray beams could only be fired when the turntable is in the flattener position" \cite{meira-goes2023cav}.
Under the nominal environment $E$, Therac-25 satisfies the above property.

\noindent
\textbf{Calculating Robustness}
We obtained $9$ feasible deviations from GPT.
Then, we computed the robustness $\Delta_f$ based on $D$. 
We found a single maximal robust deviation, showing the Therac-25 system is robust against 7 transitions.
%We obtain similar robustness analyses,  as in \cite{meira-goes2023cav}.%\textcolor{blue}{: both works consider operator errors involving mode re-selection and both identified the same non-robust transition.}

%\as{$|\Delta|=4$ was found in \cite{meira-goes2023cav}, compare to $|\Delta_f|=2 \text{ or } 1$ we retrieved.}

% For example, it is robust to a faulty beam firing, transition $('fireBeam',b, 'fireBeam')$, where the operator presses the beam button but the system does not advance to the finished state.
First, the system is robust if the operator repeatedly presses $enter$, as shown by the transitions in gray in Fig.~\ref{fig:therac}. 
% a deviation $('confirmMode', enter, 'confirmMode')$ where the enter key fails to advance the workflow, while the supervisor prevents X-ray emission until the turntable reaches the correct position.     
On the contrary, Therac-25 is not robust to mistakenly selecting the electron therapy, as depicted by the red transition in Fig.~\ref{fig:therac}. 
Any deviated plant with transition $(ch,\ e,\ ch)$ is not robust.
% The deviation $('chooseMode', e, 'chooseMode')$ models a scenario where the operator selects electron mode, but the selection is not registered; the system remains in choose mode. 
% If the operator then switches to X-ray mode, the turntable may not have time to rotate to the flattener position, potentially causing an X-ray beam to fire without proper attenuation. 
This captures the exact race condition that caused real-world Therac-25 accidents \cite{leveson1993investigation}. Similar to \cite{meira-goes2023cav}, we identified that Therac-25 is vulnerable to mode re-selection, while robust to multiple presses of $enter$.

\subsection{ABP Protocol}

\noindent
\textbf{Overview} 
The Alternate Bit Protocol (ABP) is a communication protocol that guarantees message integrity over unreliable communication channels \cite{tel2000introduction}.
We used the ABP LTS models described in \cite{zhang2020behavioral}.
The plant consists of communication channels over which messages are transmitted under normal circumstances, i.e., no loss, corruption, or duplication of messages. 
% The plant model for the ABP (Alternate Bit Protocol) system consists of two components: (1) $tx\_Channel$, a transmission channel that carries messages from sender to receiver, (2) $ack\_Channel$, an acknowledgement channel that carries acknowledgements from receiver to sender. 
% The plant $E = tx\_Channel || ack\_Channel$ represents the communication medium.
The supervisor $M_{ABP} = sender || receiver$ implements the alternating bit protocol logic. 
% The sender alternates between sending messages tagged with bit 0 and bit 1, waiting for the corresponding acknowledgement before proceeding. The receiver accepts messages and sends back acknowledgements with the appropriate bit.
The safety property ensures no repeating message bits exist, i.e., consecutive received messages must alternate between bit 0 and bit 1. 
% Formally, receiving the same bit twice in a row leads to the error state. 
% \rmg{No bit is repeated.}
% The safety property ensures correct message ordering and prevents duplication: ``Messages must be received in the order sent without duplication'', formally $G(received\_sequence\_is\_valid)$. 
% The property is encoded as an LTS with states $\{safe, waitOutput, err\}$ where two consecutive input events without an intervening output leads to the error state $err$.

\noindent
\textbf{Calculating Robustness} 
% \textit{Claude} identified 48 feasible deviations from $|(Q \times Act \times Q) \setminus R|=4*8*4-16=112$ possible transitions using constraints that channels can lose messages but cannot corrupt contents, reorder messages (FIFO property), or duplicate messages. The LLM correctly distinguished between message loss (feasible) and message corruption/reordering (infeasible). Symbolic verification found that all 48 deviations (100\%) cause violations.
% Symbolic verification found that all 48 deviations cause violations; the ABP protocol is not robust against any single feasible transition deviation. This result is significant: while the original ABP protocol is designed to handle message loss through retransmission, the feasible deviations identified by the LLM represent scenarios where the protocol's assumptions are violated.
GPT identified $|D| = 16$ feasible deviations. 
Then, we computed the robustness $\Delta_f$ based on $D$. We found two robust deviations with 8 transitions.
This deviation includes fault types such as message loss, duplication, and corruption, all of which are deviations that the ABP protocol is designed to be robust against.
Zhang et al. \cite{zhang2020behavioral} reached the similar conclusion from robustness analysis.

% \textcolor{brown}{However, due to the size of $D$ combined with the size of the ABP system, our brute-force algorithm could not calculate $\Delta_f$.
% We leave for future work implementing Algorithm~\ref{alg:rob} with heuristics to compute $\Delta_f$. }

%For example, $(('Send','Ack'), send[0], ('Send','Ack'))$ models a scenario where a $send[0]$ event occurs but the channel state does not progress, effectively a ``phantom send'' that corrupts the protocol's bit alternation. Similarly, $(('Receive','Ack'),rec[0],('Receive','Ack'))$ models a receiver that processes the same message twice, violating the no-duplication property.
%These results confirm that ABP's robustness relies critically on the FIFO and no-duplication assumptions of the underlying channel.
% \rmg{Same comment here. What is the ABP robust to? Delays, duplication, etc. THis is more interesting than just talk about the number. }

%The LLM processed 384 potential deviations and identified 42 as feasible using constraints that channels can lose messages but cannot corrupt contents, reorder messages (FIFO), or duplicate messages. The LLM correctly distinguished between message loss (feasible) and message corruption/reordering (infeasible). Symbolic verification yielded $|\Delta| = 7$ maximal robust deviations. The protocol is robust against arbitrary message losses but not against deviations that violate FIFO ordering or introduce duplication

\subsection{Results and Discussion}\label{sect:disc}
Table~\ref{tab:results} summarizes the results from both LLMs across all case studies. We determined the number of feasible transitions by manual inspection of $|D|$.
% \footnote{Due to the size of feasible deviations, we were unable to compute $\Delta_f$ for the ABP case using Claude.}.
As depicted in Table~\ref{tab:results}, both GPT and Claude identify feasible deviation sets whose cardinality is significantly smaller than that of the full transition space.
The worst reduction scenario is $67\%$, $64$ to $21$ transitions, while the best reduction scenario is $99\%$, $2048$ to $16$ transitions. ChatGPT generates smaller deviation sets, but with higher proportions of physically feasible transitions. 
For the manufacturing example, 89$\%$ of GPT-generated transitions are physically feasible, compared to 54$\%$ for Claude. 
On a more complex system like the ABP protocol, feasibility drops for both models, and the overlap between their outputs also decreases. 
Incorporating domain ontologies is a promising direction for improving LLM accuracy in future work~\cite{dean2024ontologies}.

\begin{table}[h]
\centering
\caption{Comparison of LLM-identified feasible deviations and symbolic-verified robustness. TO stands for timeout after 24 hours.}
\label{tab:results}
{\setlength\tabcolsep{4pt}
\begin{tabular}{lcccc}
\toprule
\textbf{LLM} & \textbf{Case Study} & \textbf{Mfg} & \textbf{Therac-25} & \textbf{ABP} \\
\midrule
 & $|Q_E\times Act_E\times Q_E|$ & 108 & 64 & 2048 \\
\midrule[1.3pt]
\multirow{3}{*}[-1.5ex]{ChatGPT}
  & $|D|$          & 18 & 9 & 16 \\
\cmidrule(lr){2-5}
  & Feasible       & 16 (89\%) & 6 (67\%) & 10 (63\%) \\
\cmidrule(lr){2-5}
  & $|\Delta_f|$   & 1 & 1 & 2\\
\midrule[1.3pt]
\multirow{3}{*}[-1.5ex]{Claude}
  & $|D|$          & 26 & 21 & 168 \\
\cmidrule(lr){2-5}
  & Feasible       & 14 (54\%) & 8 (38\%) & 60 (36\%) \\
\cmidrule(lr){2-5}
  & $|\Delta_f|$   & TO & 2 & TO \\
\midrule[1.3pt]
\multirow{2}{*}[-0.8ex]{In Common}
  & $|D|$          & 14 & 6 & 5 \\
\cmidrule(lr){2-5}
  & Feasible      & 13 (93\%) & 5 (83\%) & 5 (100\%) \\
\bottomrule
\end{tabular}}
\vspace{-1em}
\end{table}

Overall, although we must improve the implementation of Alg.~\ref{alg:rob}, our results demonstrate that calculating $\Delta_f$ using our neurosymbolic framework is tractable across several different case studies.
These preliminary findings indicate that our proposed framework recovers similar robustness analyses as full-space analysis on $\Delta$ while operating on a set of deviations.

\section{Conclusion}\label{sect:conclusion}

This paper presented a \textit{neurosymbolic} approach to computing robustness for discrete systems that combines LLM-based feasibility filtering with symbolic model checking.
Our framework addresses the challenges of state-space explosion and inclusion of infeasible deviations by leveraging LLMs to identify physically meaningful deviations from natural language and LTS descriptions.
Evaluation of three case studies demonstrates comparable robustness analysis results with smaller sets of deviations instead of considering all transitions.
% \rmg{effectiveness is not a good word. Say that we had a comparable robustness analysis to the cases where all transitions were used, even though a smaller set of deviations was used.}
% Notably, our analysis of Therac-25 correctly identified the race condition vulnerability that caused real-world accidents, and the ABP analysis revealed that the protocol is not robust against any single feasible deviation.
% \rmg{This should be at most three lines. Be concise on it. Recommend keeping just (1) and (2) and make it concise each}
Future work includes developing prompt engineering techniques to better identify feasible deviations, implementing interactive prompt refinement where symbolic verification results guide LLM queries to correct misclassifications. 
At the same time, analyze how well the LLMs approximate the set D. 
Moreover, we plan to investigate our framework in larger systems such as manufacturing systems. 

\bibliographystyle{IEEEtran}
\bibliography{IEEEbib}

\end{document}